\documentstyle[epsf,aps,twocolumn]{revtex}
\begin{document}

\twocolumn[
\hsize\textwidth\columnwidth\hsize\csname@twocolumnfalse\endcsname
\draft

\title{
A Vertex Correction in the Gap Equation for the
High  Temperature Superconductors }

\author{Yunkyu Bang}
\address{
Dept. of Physics, Chonnam National University,  
Kwangju 500-757, Korea}
\maketitle

\begin{abstract}
We show that 
the Migdal theorem is obviously violated in the high Tc cuprates and 
the full vertex correction should be included, in particular, in the gap 
equation,
in order to be consistent with the  anomalously strong inelastic scattering
in the region of the ``hot spots'', which is observed from the various 
normal state experiments.
The full vertex correction is obtained non-pertubatively 
by utilizing the generalized  Ward identity, 
which is shown
to hold in the important scattering channel of the pairing interaction
in the high Tc cuprates.
As a result, we find a strong enhancement of Tc from the 
vertex correction despite of the strong pair breaking effect due to the
inelastic scattering.

\end{abstract}

]

Since the discovery of the high Tc superconductors, there has been 
substantial  progress in our understanding for the large part of 
normal state anomalies, if not its microscopic origin.
The marginal Fermi liquid (MFL) ansatz\cite{MFL}, for example, captures
remarkably simple essence to  understand
the various normal state anomalies  in a unified manner.
The single and essential ingredient of the MFL  phenomenology is the assumption of the
anomalous scattering kernel only with temperature scale.
Although the  microscopic mechanism of this scattering is yet controversial, it is 
generally agreed on its existence and the essential role of it for the 
high temperature superconductivity (HTSC) phenomena.
More recently, however, it becomes clear that the MFL ansatz is rather too simple to explain
more problematic normal state anomalies, in particular, such as Hall resistance, c-axis
charge dynamics, etc.
The angle resolved photo emission (ARPES) experiment\cite{arpes}, for example, reveals
strong anisotropy of the scattering rate along the Fermi surface, conveniently termed
as the ``hot spot'' region and the ``cold spot'' region according 
to  strong or weak scattering rates in the different sections of the Fermi surface.

While it is natural to think that such an anomalous inelastic scattering, 
which is responsible for the normal state anomalies,  
should also play a role  in the superconducting 
pairing mechanism, there has been no satisfactory attempt to incorporate it in the pairing 
mechanism beyond the leading order.
Up to now, the superconducting pairing mechanisms for the high Tc cuprates 
can be largely grouped into two:
 non-Fermi liquid theories and Fermi liquid based theories.
The former is typically represented by 
the interlayer pair tunneling mechanism\cite{IPT}  
and the latter one has more variety such as  the anti-ferromagnetic paramagnon 
theory\cite{pines}, 
Van Hove singularity, or more exotic theory like various
preformed boson scenarios\cite{boson}.
Each theory has various degree of success to account the known data and we are not
in the position to judge of them.
In this paper, we would like to address the importance of the vertex correction when we
apply the conventional BCS-Eliashberg pairing approach for  HTSC.

It has been pointed out by many authors the inadequacy of the blind application of 
the Migdal theorem\cite{migdal,AFM_vertex} in HTSC. 
At the qualitative level, there are already quite simple reasons not to trust the
Migdal theorem in HTSC: 
(1) $w_D/E_F$ is  not so small ($w_D$ is the characteristic frequency of
any mediating boson); 
(2) the Fermi surface anisotropy 
(certain degree of nesting or  Van Hove singularity) poses a 
potential danger to invalidate the
phase space constraint in the Migdal theorem argument.
Nevertheless, the complexity of higher order vertex diagrams prohibits systematic progress
along this line; at most it can only indicate the danger of vertex correction by 
calculating the first order correction.

In this paper, we took rather simple short cut by observing  two experimental findings:
(1) HTSC is a d-wave pairing state mediated by strongly peaked potential in 
the momentum space, which dominantly mediates  scattering from one ``hot spot'' region to the 
other ``hot spot'' region in the Brillion zone\cite{pines};
(2) the electrons in the ``hot spot'' region has a singular self-energy correction, destroying
almost  its 
quasi-particle nature at all.
From the above two observations, we show that we are exactly opposite limit,
 for the important pairing channel,
from which the Migdal theorem is valid. In this opposite limit - we conveniently
call it the ``Ward identity limit'' - we can easily read off  the exact vertex from the
generalized Ward identity\cite{schrieffer}, given a set of reasonable assumptions.
As a demonstration, we performed a  numerical calculation to solve the model
 Tc equation for a d-wave
state with a full vertex correction. 
We find a strong enhancement of Tc
from the singular vertex correction despite of the strong pair breaking effect of the
self-energy correction.

We briefly reexamine the Migdal theorem. 
Fig.1.a shows the typical vertex correction diagram of the first order.
The typically interesting parameter regime of the momenta is that $\vec{q}, \vec{p}, 
\vec{k} \sim \vec{k}_F$
and $q_0, p_0, k_0 \sim w_D$; here and afterward
 $w_D$ is not necessarily the Debye frequency, but
some characteristic frequency of any mediating boson. 
For the Migdal limit, i.e. $\vec{q} \cdot \vec{v}_F > q_0 (\sim w_D)$, 
the correction is of the order $O(w_D/E_F)$ 
either due to the energy denominator in the electron propagators or due to
the phase space constraint.
Here the important observation is that $\vec{q} \cdot \vec{v}_F$ 
should be understood as 
$\epsilon(k+q) -\epsilon(k)$ in more general expression and particularly for the
tight binding band.
For the other limit, i.e. $\vec{q} \cdot \vec{v}_F < q_0$, 
which we call the ``Ward identity limit'', 
the correction is $O(1)$ as found  by Migdal and also more clearly demonstrated
by Engelsberg and Schrieffer \cite{schrieffer} using the generalized Ward identity.
However, the above conclusion is  true only under the following set of assumptions:
(1) electron-phonon vertex, $g(\vec{k},\vec{k'})$, has no strong momentum dependence 
-- in practice, assumed to be constant in most of analysis;
(2) similarly, the phonon dispersion, $w_q$, is assumed to be isotropic; namely, it
doesn't have any special directionality;
(3) the Fermi surface doesn't have a strong anisotropy so that it doesn't introduce
any special scattering phase space constraint; a pathological case is one dimensional
system, for example;
(4) in particular, for superconducting instability, the typical scattering process is 
involved with  large momentum exchange ($ |\vec{q}| \sim O(|\vec{k}_F|)$); for
a contrasting example,
the dynamic polarizibility, $P(\vec{q},w)$, of electron gas for $|\vec{q}|/p_F \rightarrow 0$ limit
should have an important vertex correction as a trivial violation of the Migdal 
theorem\cite{schrieffer}.

Now let us examine the situation of the high Tc cuprates.
First of all, $w_D/E_F$ is not that small since now the relevant  $w_D$ is most
probably electronic origin and also $E_F$ is renormalized to a smaller value\cite{daggotto}.
However, it is just one general fact to invalidate the Migdal theorem and it is not our
main concern in this paper.
There are two important observations just specific to the high Tc cuprates which break the 
Migdal theorem. 
First, the hole-doped high Tc cuprates is known to be a d-wave 
superconductor by now and accordingly the pairing potential $V(k,k')$ has a strong 
momentum dependence; we have in our mind  the anti-ferromagnetic spin fluctuation 
mediated potential which  is strongly
peaked for $\vec{k} - \vec{k'} = \vec{Q}$, where $\vec{Q} \simeq (\pm \pi, \pm \pi)$
\cite{pines}.
Second, the Fermi surface is anisotropic; in particular, the important parts of the Fermi 
surface with the maximum gap opened are connected by $\vec{Q}$ (see Fig.1.b).
Combining these two facts violates the Migdal assumptions (1)-(3) above.
In particular, the scattering phase space constraint - the relative probability for
satisfying $\vec{q} \cdot \vec{v}_F < w_D$ with $|\vec{q}| \sim |\vec{k_F}|$
is $O(w_D/E_F)$
with the Migdal assumptions - doesn't hold anymore.
Moreover, the last condition for superconductivity of large  momentum
exchange doesn't help, too, because the superficial Migdal condition, 
i.e., $\vec{q} \cdot \vec{v}_F > w_D$ with large $\vec{q}$ for a parabolic band becomes 
$\epsilon(\vec{k}+\vec{q})-\epsilon(\vec{k}) < w_D$ for the high Tc superconductors
when both  $\vec{k}+\vec{q}$ and $\vec{k}$ belong to the hot spot region.
To summarize, in the high Tc cuprates, because of the strongly momentum dependent pairing
potential, $V(k,k')$, inducing a d-wave pairing and the anisotropic Fermi surface from
the tight binding nature, the important scattering process for the superconducting pairing
is not in the Migdal regime ($\epsilon(\vec{k}+\vec{q})-\epsilon(\vec{k}) > w_D$) but in the 
Ward identity regime ($\epsilon(\vec{k}+\vec{q})-\epsilon(\vec{k}) < w_D$).
In the following, we show that we can indeed extract the full vertex correction 
of this scattering
channel ($\vec{k}, \vec{k}+\vec{q} \in$ ``hot spot'' region) from the generalized 
Ward identity \cite{pertubation}.

Engelsberg and  Schrieffer\cite{schrieffer}
derived the following general identity from the particle 
number conservation condition.

\begin{equation}
q_0 \Gamma^{0}(k;k+q) - \vec{q} \cdot \vec{\Gamma}(k; k+q) = G^{-1}(k+q) - G^{-1}(k)
\end{equation}
\noindent
where $k, q$ stands for the four momenta, i.e., $k = (k_0, \vec{k})$, and
$\Gamma^{0}$ and $\vec{\Gamma}$ are scalar and vector vertices, respectively.
$G(k)=(k_0 - \epsilon(k) -\Sigma(k))^{-1}$ is the full electron green function 
with a self-energy.
The above equation is exact for any $k$ and $q$ and the famous ``Ward identity'' is
derived by taking $\vec{q}=0$ and $q_0 \rightarrow 0$ limit, so that
$\Gamma^{0}(k;k) = 1 - \partial{\Sigma(k)}/\partial{k_0}$. 
The other limit, $q_0=0$ and $\vec{q} \rightarrow 0$, can be used to obtain the
vector vertex $\vec{\Gamma}(k;k)$.
Since we assume a d-wave pairing in the high Tc superconductors and the maximum gap is formed
around the ``hot spots'', the important part of the pair potential is the process 
which scatters from one ``hot spot'' region to the other ``hot spot'' region.
Consequently, our interest is the scalar vertex, $\Gamma^{0}(\vec{k},k_0;\vec{q} \sim \vec{Q},
q_0 < w_D)$, $\vec{k} \in$ ``hot spot'' region. 
In general, there is no simple solution for this unless we have a simple form of
$\Sigma(k)$.
Theoretical\cite{chubukov} and experimental\cite{arpes} studies show that the self-energy
correction around the ``hot spots'' is singular in frequency but
its momentum dependence is negligible. 
Therefore, we can separate the scalar and vector vertices to a good degree of approximation
as follows,
for the interesting regime of 
$\vec{q} \sim \vec{Q}$, $q_0 < w_D$ and $\vec{k} \in$ ``hot spot'' region\cite{hot_spot}.

\begin{eqnarray}
q_0 \Gamma^{0}(\vec{k},k_0;\vec{Q},q_0) & \cong & \\ \nonumber
q_0 -[(\Sigma(\vec{k}+\vec{Q},k_0+q_0) - \Sigma(\vec{k},k_0))], & &\\ 
\vec{Q} \cdot \vec{\Gamma}(\vec{k},k_0;\vec{k}+\vec{Q},k_0+q_0) & \cong &
\epsilon(\vec{k}+\vec{Q}) - \epsilon(\vec{k}),  
\end{eqnarray}
\noindent
where $\epsilon(\vec{k})$ is the bare band dispersion.
As argued above, we assumed that $\Sigma(\vec{k}+\vec{Q},k_0) \simeq \Sigma(\vec{k},k_0)$
when both $\vec{k}+\vec{Q}$ and $\vec{k}$ belong to the hot spot region for the above
separation.

Now to be specific, we approximate the self-energy for the hot spot region as the
marginal Fermi liquid type as follows.
\begin{equation}
Im \Sigma(w) = \alpha |w| ~~ for ~~ |w| < w_c,
\end{equation}
\noindent
where $w_c$ is the high energy cutoff,
and $\alpha$ is about 0.6 from experiment\cite{schlesinger}
but it can be treated as a parameter for our purpose.
In reality, the self-energy correction can be even more singular in the hot spot region
\cite{chubukov},
or maybe just Fermi liquid type ($\Sigma^{''} \sim w^2$) for
very low frequency limit.
In any case, our main conclusion doesn't change, i.e. 
{\em large vertex correction 
from the anomalous scattering in the hot spot region.} 
Here the marginal Fermi liquid type self-energy assumption
is just for demonstration purpose; 
nevertheless we think it is 
still reasonable assumption in view of  experiments.
Now the full self-energy in the Matsubara frequency is written as 
$\Sigma(w_n) = - iw_n \frac{\alpha}{\pi} ln \frac{w_n ^2 + w_c ^2}{w_n ^2}$.
And the corresponding scalar vertex is written as
\begin{eqnarray}
\Gamma^0(\vec{k}, w_n;\vec{q} \sim \vec{Q},  \Omega_n)&  = &\\ \nonumber
1 -[\Sigma(w_n+\Omega_n) - \Sigma(w_n)]/ i \Omega_n, & &
\end{eqnarray}
where $w_n= \pi k_B T (2n +1)$ and $\Omega_n = \pi k_B T (2n)$ are 
the fermionic and bosonic Matsubara frequencies, respectively.
Since we are interested in $\Omega_n < w_D$ regime, we take the $\Omega_n \rightarrow 0$
limit from the above equation. Now the full vertex for $\vec{q} \sim \vec{Q}$, 
$\vec{k} \in$ ``hot spot''  region
in the static limit is the following.
\begin{eqnarray}
\Gamma^0(w_n) & = & [1 + \frac{\alpha}{\pi} ln \frac{w_n ^2 + w_c ^2}{w_n ^2} ] \\ \nonumber
& & -2 \frac{\alpha}{\pi} \frac{w_c ^2}{w_n ^2 + w_c ^2}.
\end{eqnarray}
\noindent
This vertex shows the $ln T $ divergence when $w_n \rightarrow 0$ limit due to the singular
self-energy\cite{FL}. 
The first term of the Eq.(6) is nothing but $Z_n$ in Matsubara formalism; 
$Z_n$ is defined as $iw_n - \Sigma(w_n) = iw_n Z_n$.
We notice that the singular suppression of the effective coupling constant
due to $Z_n$ as 
$g_{eff}=g/Z_{n}$ can be almost recovered now as $g_{eff}=g \Gamma^0_{n}/Z_{n}$
by the corresponding vertex correction reminiscent of the original Ward identity in QED.
With this vertex we solve the model Tc equation for a d-wave state.
The Tc equation is written as
\begin{equation}
\Delta(k) = - k_B T \sum_{w_n} \sum_{k'}\frac{ V(\vec{k},\vec{k'}) \Gamma^0(w_n)^2
\Phi(\vec{k},\vec{k'})}
{Z_n^2 w_n ^2 + \xi_{k'} ^2} \Delta(k').
\end{equation}

Some remarks are in order about the above Tc equation. First of all,
the above Tc equation is the static limit of the Eliashberg equation, so that it would
be the BCS equation were it not for the self-energy correction, $Z_n$, and the vertex 
correction, $\Gamma^0(w_n)$ of static limit. 
We could have solved
a full Eliashberg equation with a given dynamic pair potential and it would be no problem
to include the dynamic vertex correction,  $\Gamma^0(w_n;\Omega_n)$, from Eq.(5). 
We think that qualitative results would not
be changed.
Second, the vertex correction we put in the Tc equation is valid only when both  
incoming momentum, $\vec{k}$, and  outgoing momentum, $ \vec{k'}=\vec{k} + \vec{Q} +\Delta k $,
belong to the hot spot region. 
In order to simulate this hot spot scattering channel constraint, we introduced
a function
$\Phi(\vec{k},\vec{k'})$, which is just 1 only when
both $\vec{k}$ and $\vec{k'}$ belong to
the hot spot region connected by $\vec{Q}$; otherwise it 
replaces $\Gamma^0(w_n)$ with the bare vertex, $\Gamma^{0}=1$.
It means that we do not consider any vertex correction for other scattering channels
other than this special scattering channel. Some justifications for this are: (1) for 
other scattering channels, simply the vertex correction should not be as singular as 
for the hot spot scattering channel; (2) even if there should be some vertex correction,
its effect in the Tc equation is suppressed by the pair potential, 
$V(\vec{k},\vec{k'})$,
since it is peaked only for the hot spot scattering channel.

For simplicity of  numerical calculation, we assume 
$V(k,k') = V |sin(\phi - \phi')|$ and $\Delta(k) = \Delta_{max} cos (2 \phi)$
in  2-dimensional momentum space, a circular Fermi surface, and $\phi$ is the angle along the 
Fermi surface. For all calculations, we take $w_c=0.5 eV$, $w_D=0.3 eV$ ($w_D$ enters
as the BCS cutoff in the momentum summation)  and 
the coupling constant parameter $\lambda = V N(0)$ is taken to be 1.5
($N(0)$ is the density of states at the Fermi level).
This choice of parameters is only for exemplary purpose.
Also for the hot spot scattering channel constraint function, 
$\Phi(\vec{k},\vec{k'})$,
we define the the hot spot region by the angle $\theta_{hot}$ as
indicated in Fig.1b;
now  $\Phi(\phi,\phi')$=1 only when $\phi$, $\phi^{'} \in$ ``hot spot regions'',
otherwise it will set $\Gamma^0(w_n)=1$.

In Fig2, we show  Tc as a function of the strength of the self-energy 
correction, $\alpha$, for different values of $\theta_{hot}$.
For example, $\theta_{hot}= \pi/2$ is the case that the whole 
Fermi surface is treated
as the hot spot region. As expected, the  $\theta_{hot}= \pi/2$ case (open square)
shows an extreme slow-down of the suppression rate of Tc
because of  the over-imposed vertex correction; the solid square indicates the result
with no vertex correction at all
 but only with the self-energy correction, showing rapid
suppression of Tc with increasing $\alpha$. 
In between these two curves, the results with
$\theta_{hot}=\pi/4$ and $\pi/8$ are shown and these should be the more realistic cases.
In short, Fig.2 shows the dramatic effect of the vertex correction for Tc
even with a very narrow area of the hot spot region (see $\theta_{hot}=\pi/8$ case).
When considering $\alpha \simeq 0.6$ for YBCO (Tc $\sim$ 90 K)\cite{schlesinger},
without the vertex correction at all (solid square), Tc $\simeq$ 24 K only;
the set of parameters (the dimensionless coupling constant, $\lambda=1.5$,
and a characteristic energy of mediating boson, $w_D=0.3 eV$ ) is already 
quite favorable choice for the pairing. 
Only with $\theta_{hot}=\pi/8$, 
Tc is enhanced
more than by a factor of 3 (Tc $\simeq$ 85 K) for the same $\alpha =0.6$.
The message is that in order to achieve Tc $\sim$ 100 K of a  d-wave state in HTSC,
with strong inelastic scattering but without including
the corresponding vertex correction, we are
forced to choose rather unrealistic parameters in the  conventional
pairing model.

In conclusion, we show that 
{\em the Migdal theorem is
maximally violated in  HTSC, not only because of $w_D/E_F \sim O(1)$ but
more importantly because of the strongly momentum dependent pairing potential,
$V(\vec{k},\vec{k'})$, and the existence of hot spot region in the anisotropic Fermi surface},
which are dominantly scattered by the pairing potential
in the main pairing channel of a d-wave state.
Then, we show that we are in the ``Ward identity limit'' which allows us to
extract the full vertex correction from the self-energy.
Assuming a phenomenological self-energy ($\Sigma^{''}=\alpha |w|$) and
taking into account the corresponding vertex correction 
around the ``hot spots'',
we solved the model Tc equation
for a d-wave pairing including both the self-energy correction and the vertex 
correction.
The results show the dramatic enhancing effect of Tc by the vertex correction despite
of the strong pair breaking self-energy correction. Considering the problem of
choosing realistic values of  parameters for a d-wave pairing model, our
result enforces the essential role of the singular vertex correction, at least
for a conventional BCS-Eliashberg type pairing scenario.
Amusing observation is that the strong inelastic scattering in the high Tc cuprates,
which is the key entity responsible for the various anomalous normal state properties\cite{MFL},
turns out to be not so much destructive 
for superconductivity in the end,
in contrast to the conventional superconductivity, thanks to the singular vertex correction.
The origin of it is the strong anisotropies 
in the pairing potential $V(\vec{k},\vec{k'})$, the Fermi surface, and the order
parameters $\Delta(\vec{k})$, which are all related, though.
Therefore, in HTSC the strong correlations not only in frequency domain
but also in spatial domain play  important roles on their own to make all
these unusual materials.

We acknowledge the financial support of the Korea Research Foundation, 1996, 
the Korean Science and Engineering Foundation (KOSEF) 
Grant No. 961-0208-048-1, 976-0200-006-2,
and the  KOSEF through the SRC program of SNU-CTP.

\begin{figure}
\epsfxsize=8.5cm
$$\epsffile{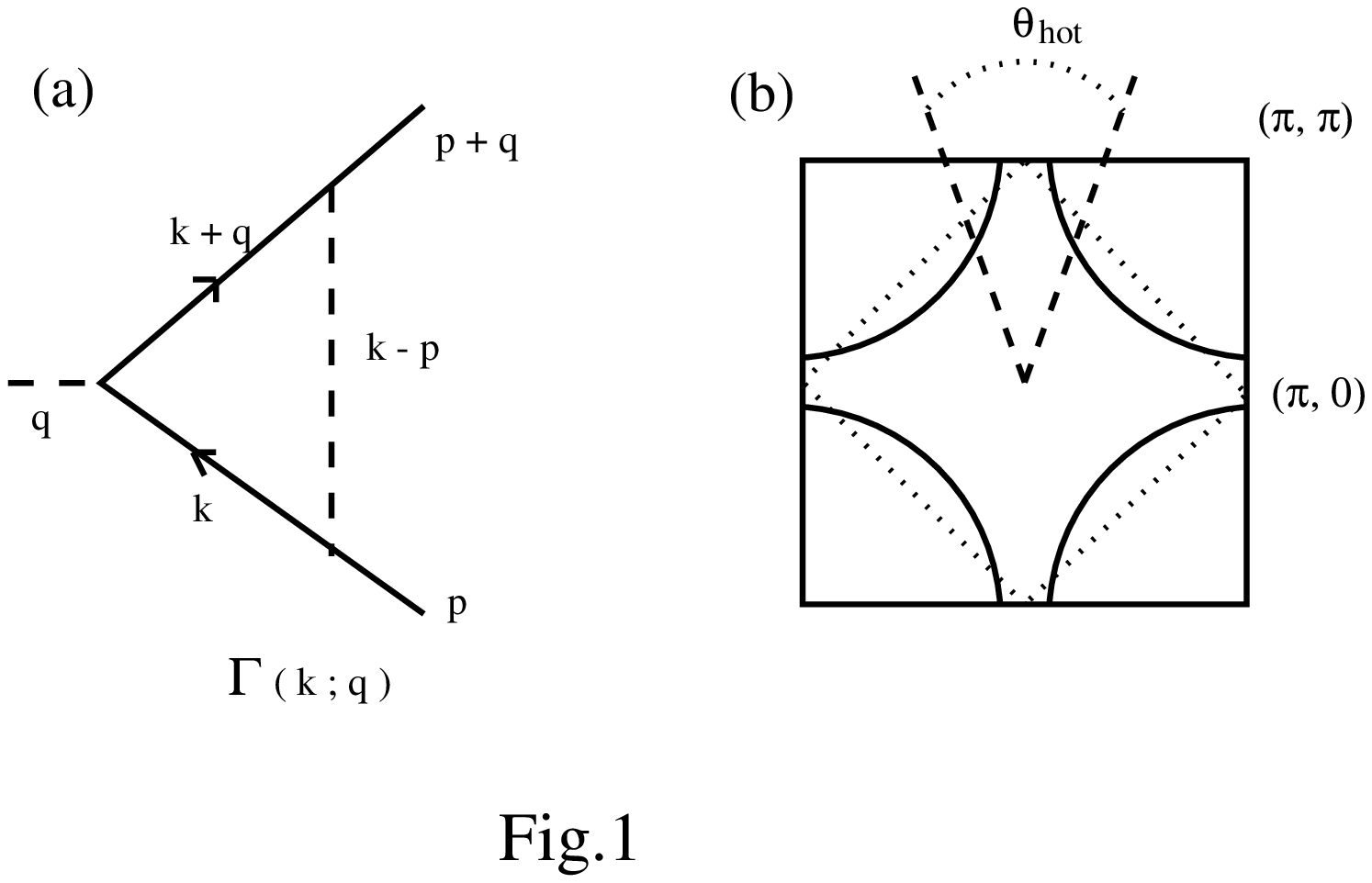}$$
\caption{ (a) Vertex diagram of the first order; solid lines
are electron propagators and dotted line is any bosonic propagator.
$k$ and $q$ stand for four momenta, i.e., $k=(k_0, \vec{k})$.
(b) Schematic Fermi surface of HTSC; the areas around zone corners,
($\pi,0$), and its symmetry related points, are the
``hot spot'' region, which are connected by the momentum $\tilde{\vec{Q}} \simeq
(\pm \pi,\pm \pi)$.
The angle $\theta_{hot}$ defines the hot spot region 
quantitatively for the numerical calculation.}
\label{fig1}
\end{figure}

\begin{figure}
\epsfxsize=8.5cm
$$\epsffile{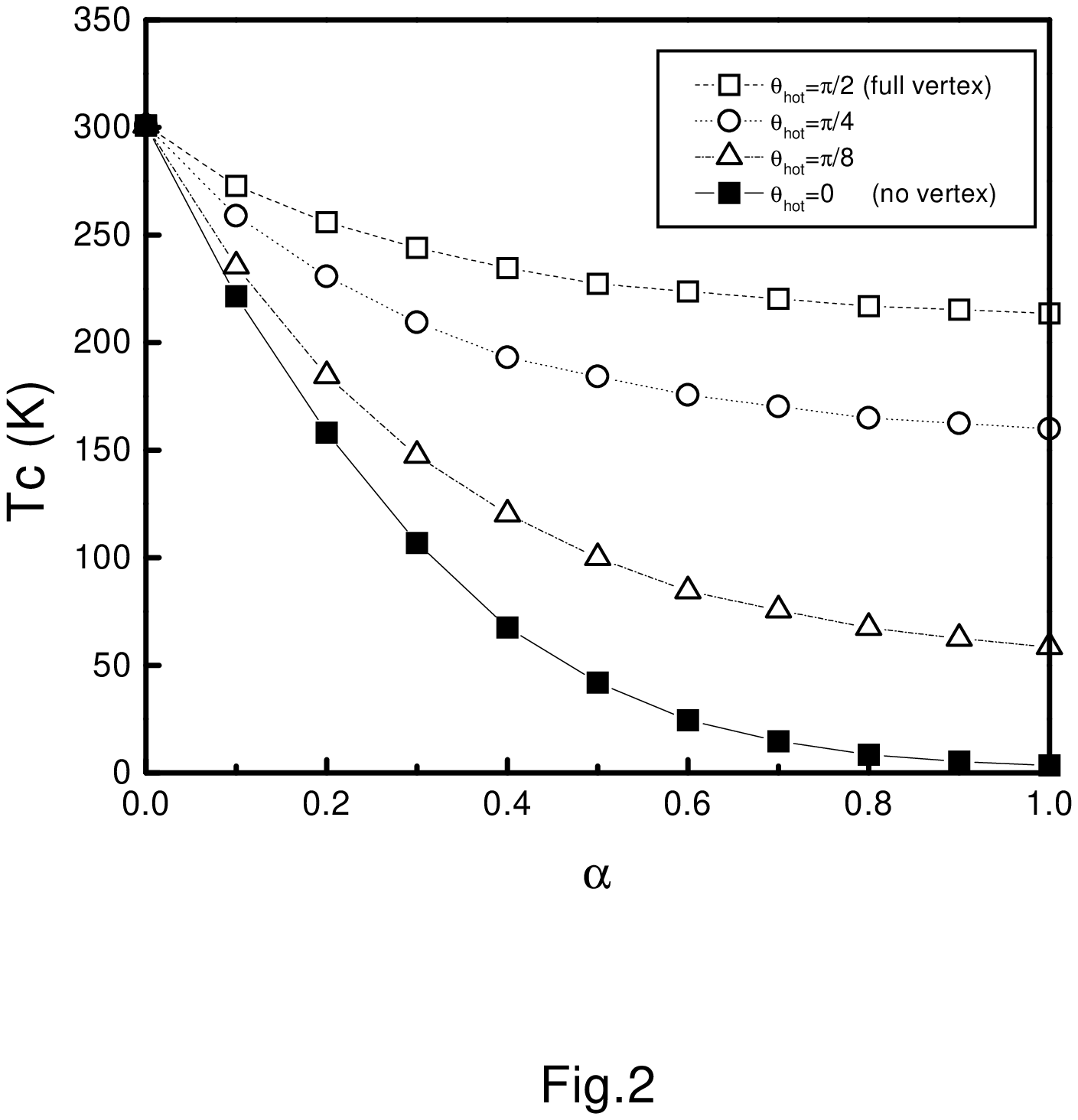}$$
\vspace{-5cm}
\caption{Tc vs $\alpha$, the pair breaking parameter defined in 
$\Sigma^{''}=\alpha |w|$; solid square is the one without vertex correction,
and 
open upper triangle, open circle, and open square
are the ones with more vertex correction (wider area of the hot spot region)
in increasing order.}
\label{fig2}
\end{figure}


\begin{references}
\bibitem[12]{hot_spot}
The hot spot region is defined as $\Delta k \xi < 1$. $\Delta k$ is the distance
of $k$ from the nearest ``hot spot'' along the Fermi surface. In optimally doped 
YBCO sample, $\xi \sim 2$ above $Tc$, therefore the hot spot region can have substantial 
fraction along the FS (see, e.g., B. P. Stojkovic and D. Pine in Ref.[11])
\bibitem[14]{FL} If a Fermi liquid like self-energy, say, 
$\Sigma^{''}=\tilde{\alpha} w^2$ with $\tilde{\alpha} \sim O(1/E_F)$, the vertex 
correction is still  O(1) as $w_n \rightarrow 0$.
\end{references}
\end{document}